\documentstyle[12pt,epsfig]{article}
\def\ra{\rightarrow}
\def\ben{\begin{subequations}}
\def\be{\begin{equation}}
\def\een{\end{subequations}}
\def\ee{\end{equation}}
\def\beq{\begin{eqalignno}}
\def\eeq{\end{eqalignno}}
\def\bea{\begin{eqnarray}}
\def\eea{\end{eqnarray}}

\def\f2gam{\mbox{$ F_2^\gamma $}}

\newcommand{\ea}{\end{array}}
\def\sd{\strut\displaystyle}
\newcommand{\PL}[3]{{ Phys. Lett.}       {\bf #1} {(19#2)} {#3}}
\newcommand{\PRL}[3]{{ Phys. Rev. Lett.} {\bf #1} {(19#2)} {#3}}
\newcommand{\PR}[3]{{ Phys. Rev.}        {\bf #1} {(19#2)} {#3}}
\newcommand{\ZP}[3]{{ Z.  Phys.}         {\bf #1} {(19#2)} {#3}}
\newcommand{\NP}[3]{{ Nucl. Phys.}       {\bf #1} {(19#2)} {#3}}
\title{ 
  {\large \bf  Hadronic Total Cross-sections Through Soft Gluon Summation  in 
Impact Parameter Space}
}
\author{ A. Grau$^1$, G. Pancheri$^2$, Y.N. Srivastava$^3$\\
{\small \it ${}^{1)}$Departamento de F\'\i sica Te\'orica y del Cosmos,
Universidad de Granada, Spain}  
 \\
{\small \it ${}^{2)}$INFN, Laboratori Nazionali di Frascati, P.O. Box 13,
I-00044 Frascati, Italy}\\
{\small \it ${}^{3)}$Physics Department and INFN, University of Perugia, Perugia, 
Italy}
} 
\begin{document}
\input FEYNMAN
\maketitle
\begin{abstract}

The Bloch-Nordsieck  model for the parton distribution of hadrons
 in impact parameter space, 
constructed using soft gluon summation, is investigated in detail.
 Its dependence upon the infrared structure of
the strong coupling constant $\alpha_s$ is discussed,   both for
 finite as well as  singular, but integrable, $\alpha_s$. 
The formalism is applied to the prediction of total 
proton-proton and proton-antiproton cross-sections, where screening,
due to
soft gluon emission from the
initial valence quarks, becomes evident.
\end{abstract}

\section{Introduction}

In this paper we address some phenomenological
implications of the infrared behaviour of the 
strong coupling constant $\alpha_s$\cite{yuri}. 
In particular, we examine some models for the total
proton-proton and proton-antiproton cross-sections and show the
dependence of the rise with energy of the cross-section upon the small 
$k_t$ behaviour
of $\alpha_s$, through the mechanism of soft gluon
summation. 
In a previous paper \cite{bn}, soft gluon summation techniques
have been applied to develop a model for the impact parameter distribution
of partons in hadronic collisions. 
According to this model, the distribution in impact parameter space
 ($b$-distribution) is  the Fourier transform 
of the transverse momentum 
distribution of the colliding parton pair,  and is obtained by using the
 Bloch-Norsdieck technique for soft gluon summation, developed some
time ago to 
describe hadronic
transverse momentum distributions \cite{PSS,DDT,PP,GRECO}.
This model for the $b-$distribution of partons is used in the context of
 eikonal models for total cross-sections, and in
particular in the context of the eikonal mini-jet models, where the rise
with energy is driven by the jet cross-section calculated from 
QCD. In order
to make full use of QCD for this particular problem,
it is necessary that not
only the energy dependence be derived from QCD, but also
the $b-$dependence, at least for what concerns the hard part of the
cross-section : it may otherwise be possible to obscure the difficulties 
of QCD inspired models
through various parameters which are still present in it. One of
the difficulties is that the QCD cross-section rises too fast
with energy to be able to accomodate both the early rise (around $\sqrt{s}
=10\div 20\ GeV$) and the high energy behaviour at
 $\sqrt{s}\ge 200\div300\ GeV $ and beyond. In some mini-jet models
 the too abrupt rise  of the mini-jet
cross-section is softened by modifying the small x-behaviour of the
parton densities. Our alternative proposal, discussed in detail in this paper, 
is to   regulate 
the rise of the cross-section through
soft gluon emission.

In Sect. 2 we present a brief description of
the eikonal mini-jet
model.
In Sects.3$\div$  6 we shall analyze the structure of the
 Block-Nordsieck model for the $b-$distribution of partons, first recalling
 the main features of the model, and then studying ,
 analytically as well as numerically, its behaviour employing various 
phenomenological models for the $k_t \ra 0$ behaviour of the strong coupling
constant $\alpha_s$.  In all cases, we shall compare our results 
with those from a model in which the
matter distribution of partons is obtained from the electromagnetic form
factor of the colliding hadrons. In the last two sections, Sects. 7 and 8,
 we shall
study the predictions of  the Bloch-Nordsieck model for total
cross-sections and shall compare our results for proton-proton
and proton-antiproton collisions with other models and present data.
It will be shown that
 the model,
with a singular but otherwise integrable behaviour of $\alpha_s$, is 
flexible enough to
accomodate both the early rise with energy as well as  present data from the
Tevatron.

\section{Eikonal mini-jet model for total cross-sections}

Ever since the first observation of the rise of proton-proton 
total cross-section, the
suggestion was advanced that such rise was due to the increasing
 (with  energy) number of
hard collisions taking place among the hadron
constituents \cite{CLINE}. This ans\"atz was subsequently quantified by the
mini-jet model, which proposes to calculate the total inelastic cross-section
from the jet cross-section obtained from QCD \cite{GAISSER,PS}.
The unitarized version of the mini-jet model is represented by
the eikonalized minijet model\cite{DURAND,CAPELLA,BLOCK},
 in which the total cross-section 
is given by
\be
\label{sigtot}
\sigma_{tot}=2\int d^2{\vec b}\left[1-e^{-n(b,s)/2}\right]
\ee
with
\be
\label{nbs}
n(b,s)=A(b)[\sigma_{soft}+\sigma_{jet}]
\ee
and $A(b)$ a function which represents the impact parameter distribution
of partons in the collision. In its most intuitive formulation, the
overlap is obtained from the Fourier transform of the electromagnetic
form factors ${\cal F}_1$ and ${\cal F}_2$ of the colliding hadrons, i.e.
\be
\label{aff}
A_{FF}(b)={{1}\over{(2\pi)^2}}\int d^2{\vec q} e^{ib\cdot q} {\cal F}_1(q)
{\cal F}_2(q)
\ee
The model which uses this  overlap function, hereafter called the form factor 
(FF)
model, although attractive, is of course not parameter free, as it 
depends on the scale parameters characterizing the form factors.

The two cross-section $\sigma_{soft}$ and $\sigma_{jet}$ are respectively
a non-perturbative term and a function of energy obtained by integrating the
QCD jet cross-section from a minimum $p_t$ value, $p_{tmin}$, to the maximum
kinematically allowed. This quantity increases with energy at fixed $p_{tmin}$,
depending upon various QCD controlled quantities like the parton densities,
in particular, and very strongly, upon the small x-behaviour of the gluon 
densities. In fact, the kinematic lower limit in the x-integration for
the jet cross-section is
given by $x_{min}=4p^2_{tmin}/s$, and it can be as low as $10^{-6}$ at
 Tevatron 
energies. With such small x-values, the jet cross-section grows much too rapidly
as $s$ increases and so does the eikonalized cross-section.
In order to apply the mini-jet model to data, a screening effect is
obtained  either using the much less 
dangerous limit $\sqrt{x_{min}}$ or  softening the small-x singularity with a
cutoff parameter.
In this way, the above model
can reproduce the energy rise, but with some further
 modifications,
notably in $A(b)$. In particular, in order to obtain reasonable agreement with
the data it is also necessary to modify the simple form factor model, by
 allowing for
 different values of the scale parameters for the low and high energy region.

Our approach is different. We believe that the function $A(b)$ is not
 a constant in energy and  for the hard part of the collisions
we have proposed a model in
which soft gluon emission is responsible for
the b-distribution of the colliding partons. Since the overall
soft gluon emission summation is energy dependent, we expect such model can
modify and complement the mini-jet model description of total cross-sections.

\section{Bloch-Nordsieck formalism in impact parameter space}

The Bloch-Nordsieck distribution depends upon the energies of the
colliding quarks and gluons and is thus, although mildly, energy dependent. 
In this section we shall recapitulate the main features of this model, whose
general structure was derived in ref.\cite{bn}.
As described, our proposed impact parameter space
 distribution for
a pair of partons i and j is given by
\be
\label{ourAB}
A_{BN}= {{e^{-h(b;M,\Lambda)}}\over
{2\pi \int bdb 
 e^{-h(b;M,\Lambda)}}}
\end{equation}
 where 
\begin{equation}
\label{hdb}
h(b;M,\Lambda)={{2 c_{ij}}\over{\pi}}\int_0^M {{dk_t}
\over{k_t}}\alpha_s(k^2_t)
\ln{{M+\sqrt{M^2-k_t^2}}\over{M-\sqrt{M^2-k_t^2}}}
[1-J_0(k_t b)]
\end{equation}
with $c_{ij}=4/3$ for a quark-antiquark pair.
In eqs.(\ref{ourAB},\ref{hdb}) the hadronic scale M
 accounts for the maximum energy allowed to each single soft gluon emitted in 
the
collision. This quantity depends upon the energy of the
colliding parton pair and, through this, upon the energy of
the initial colliding hadrons.  The main point of our model is that  soft
 gluon emission destroys the collinearity of the
colliding partons.  Let  us distinguish now
between valence partons and gluons or sea quarks. In first approximation, 
gluons and sea quarks can be considered as having the same non-collinearity as 
the initial
valence quarks  which emit them during the hadronic collision (a different case
will be that of the photons, which we shall discuss in a different paper). 
To leading order we can now assume that the  impact parameter distribution of
all type of parton pairs is the same as that of the valence quarks. This
approximation is in the same spirit as the one for which the impact 
parameter distribution in the form factor model is given by the
Fourier transform of the electromagnetic form factors, i.e. matter
 distribution  follows charge distribution.

In the calculation of total cross-sections with the eikonalized mini-jet model,
 the distribution (\ref{ourAB})
 appears
convoluted with parton densities and jet cross-sections. In ref.\cite{bn},
we proposed to write the average number of collisions at impact parameter $b$ 
as 
\be
\label{nbimp}
n(b,s)=n_{soft}(b,s)+\sum_{i,j,}\int{{dx_1}\over{x_1}}
\int{{dx_2}\over{x_2}} f_i(x_1)f_j(x_2)\int dz \int dp_t^2
A_{BN}(b,q_{max}){{d\sigma}\over{dp_t^2 dz}}
\ee
where $f_i$ are the quark densities
 in the colliding hadrons, $q_{max}$ is the maximum transverse momentum 
allowed by kinematics to a single gluon emitted by the initial $q\bar q$ pair, 
  $z={\hat s_{jet}}/(sx_1x_2)$,  and ${{d\sigma}\over{dp_t^2 dz}}$ is the
  differential cross-section for process 
\be
\label{process}
q\bar q \to jet\ jet + X
\ee
for a given $p_t$ of the  produced jets with c.m. energy 
$\sqrt{{\hat s}_{jet}}$. The jet pair   in process 
(\ref{process}) is the one produced through any subprocess
initiated by the valence quark-antipark pair, thus 
it could be gluon jets, or quark jets. In Fig. 1 we show some typical 
subprocesses which contribute to (\ref{process}).
\begin{figure}
\textheight 800pt 
\textwidth 450pt
\begin{picture}(10000,18000)
\drawline\fermion[\E\REG](0,18000)[5000]
\put(0,19000){valence quark }
\drawline\fermion[\E\REG](\pbackx,\pbacky)[8000]
\drawline\gluon[\SE\REG](\pfrontx,\pfronty)[4]
\drawline\gluon[\E\REG](\pbackx,\pbacky)[5]
\put(\pbackx,\pbacky){jet}
\drawline\gluon[\S\REG](\pfrontx,\pfronty)[5]
\drawline\gluon[\E\REG](\pbackx,\pbacky)[5]
\put(\pbackx,\pbacky){jet}
\drawline\gluon[\SW\REG](\pfrontx,\pfronty)[4]
\drawline\fermion[\W\REG](\pbackx,\pbacky)[5000]
\put(0,2000){valence quark}
\drawline\fermion[\E\REG](\pfrontx,\pfronty)[8000]
\put(19000,19000){valence quark}
\drawline\fermion[\E\REG](20000,18000)[5000]
\drawline\fermion[\E\REG](\pbackx,\pbacky)[8000]
\drawline\gluon[\SE\REG](\pfrontx,\pfronty)[3]
\drawline\fermion[\E\REG](\pbackx,\pbacky)[5000]
\drawline\fermion[\SE\REG](\pfrontx,\pfronty)[3000]
\drawline\fermion[\E\REG](\pbackx,\pbacky)[5000]
\put(\pbackx,\pbacky){jet}
\drawline\gluon[\S\REG](\pfrontx,\pfronty)[3]
\drawline\fermion[\E\REG](\pbackx,\pbacky)[5000]
\put(\pbackx,\pbacky){jet}
\drawline\fermion[\SW\REG](\pfrontx,\pfronty)[3000]
\drawline\fermion[\E\REG](\pbackx,\pbacky)[5000]
\drawline\gluon[\SW\REG](\pfrontx,\pfronty)[3]
\drawline\fermion[\E\REG](\pbackx,\pbacky)[5000]
\drawline\fermion[\W\REG](\pfrontx,\pfronty)[5000]
\put(20000,2000){valence quark}
\end{picture}
\caption{Two  typical subprocesses contributing to the rise of
the total proton-proton cross-section}
\end{figure}
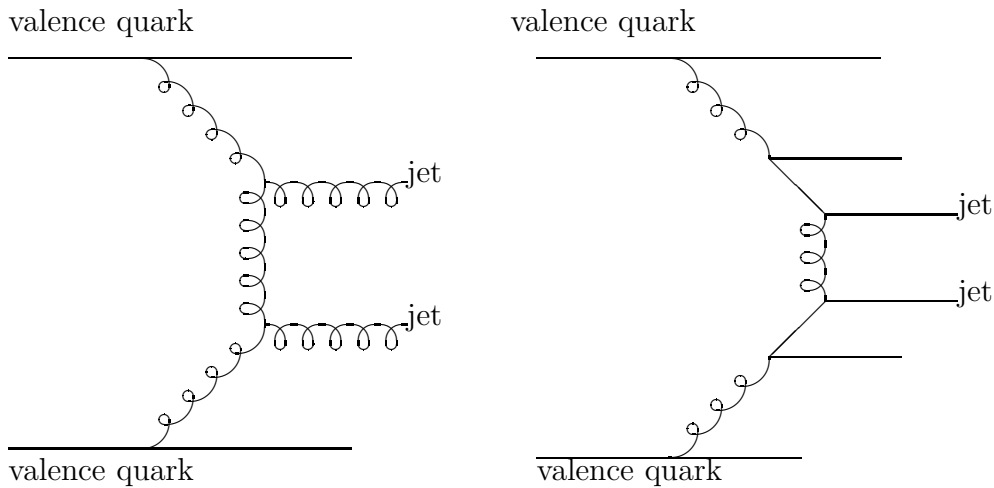
For high energy and low $p_t$,
most of the jets are  produced through scattering of
gluons emitted by a valence  quark 
pair which continues
undetected after  emission.
In principle, an exact calculation of this model for
$n(b,s)$ would require to know 
 $A(b,q_{max})$ for each subenergy $\hat s$ of the quark-antiquark pair because 
for process (\ref{process}) 
\be
\label{qmaxjets}
q_{max}({\hat s})={{ \sqrt{{\hat s}} }\over{2}}
 (1-{{{\hat s_{jet}}}\over{{\hat s}}})
\ee
and then one would need to calculate $n(b,s)$ for
each s value, through convolution for all parton densities
and all subprocesses. This procedure is at present unpractical for this
problem, since the $b$-parameter dependence applies to the
initial valence pair. What is available, through various 
parametrizations,  is parton densities after $Q^2$ evolution, 
for all type of partons, whereas the 
above formulation would require to apply corrections and evolution
in expressions which depend upon the impact variable $b$.
In any case, before recommending to embarque in such a time-consuming 
integration, 
one can  study the properties of  the proposed model,
  adopting 
some approximations, which allow for phenomenological calculations.
The approximation  described in \cite{bn} is
\be
\label{avn}
n(b,s)=n_{soft}+A_{BN}\sigma_{jet}(s,p_{tmin})
\ee
where
$A_{BN}$ is the function $A_{BN}(b,<q_{max}>)$ evaluated at the
 value $M=<q_{max}>$,  obtained by
averaging over all parton
densities and jet subprocesses. In the next section we shall 
evaluate $<q_{max}>$ for different energies of the colliding hadrons
and for different $p_{tmin}$ values.

\section{The scale dependence : $q_{max}$}
\begin{figure}[htbp]
\begin{center}
\mbox{\epsfig{file=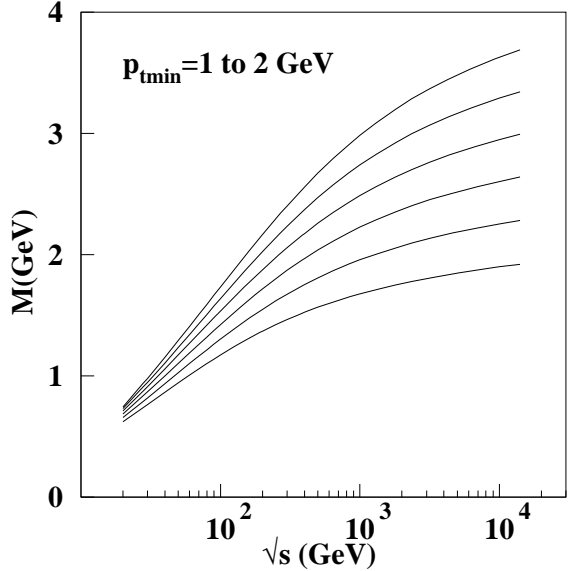,height=6.cm}}
\end{center}
\vspace{1cm}
\caption{The maximum value kinematically allowed for the transverse 
momentum of the single gluon, averaged over densities (GRV-LO 
parametrization) and for different $p_{tmin}$ values, as a function of the 
c.m. energy of the colliding protons.}
\label{qmaxfig}
\end{figure}

Using the expression
\be
\label{qmaxav}
M\equiv <q_{max}(s)>={{\sqrt{s}} 
\over{2}}{{ \sum_{i,j}\int {{dx_1}\over{ x_1}}
f_{i/a}(x_1)\int {{dx_2}\over{x_2}}f_{j/b}(x_2)\sqrt{x_1x_2} \int dz (1 - z)}
\over{\sum_{i,j}\int {dx_1\over x_1}
f_{i/a}(x_1)\int {{dx_2}\over{x_2}}f_{j/b}(x_2) \int(dz)}}
\ee
with $z_{min}=4p_{tmin}^2/(sx_1x_2)$, one can plot
the quantity $M$ as a function of $\sqrt{s}$ for different values 
of $p_{tmin}$. This
is shown in Fig.\ref{qmaxfig}, where we have used GRV(LO)
 \cite{GRV} parton densities for
 proton proton collisions.

One sees that, for $\sqrt{s}\approx 50 \div 10^4\ GeV$, 
 the  range of values for $M$ is between 
0.5 and 4 GeV for $p_{tmin}=1 
\div 2\ GeV$. For these typical values, one can now calculate 
 $h(b;M,\Lambda)$ and subsequently $A(b,M)$. Our point of interest in this
 paper is also to relate the rate of rise of the total cross-section with
the behaviour of $\alpha_s$ in the infrared region. The stronger 
the singularity as $k_t \rightarrow 0$, the larger $h(b;M,\Lambda)$, the
faster $A_{BN}$ goes to zero and the stronger will be the suppression
produced by soft gluon emission.
We shall
now quantify this statement with numerical calculations.

\section{$\alpha_s$ dependence in the function $h(b;M,\Lambda)$}

We start by showing how the b-dependence of  $h(b;M,\Lambda)$ 
varies according to 
 the behaviour of $\alpha_s$ in the very
low $k_t$ region. Because of the many uncertainties we shall
work with the one-loop expression for $\alpha_s$ and
 shall use two different models, each of them 
characterized
by a set of parameters, i.e. the frozen $\alpha_s$ model
used in
\cite{HMS,ALTARELLI}
where 
\begin{equation}
\label{altarelli}
\alpha_s(k_t^2)={{12 \pi}\over{33-2 N_f}} {{1}\over{\ln[(k_t^2+a^2 
\Lambda^2)/\Lambda^2)]}}
\end{equation}
which depends upon the  parameter set  \{$\Lambda,a$\} 
and in which 
$\alpha_s$
goes to a constant value as $k_t$ goes to zero. An altogether different
 model
is the singular $\alpha_s$ model, described in \cite{bn}
with
\begin{equation}
\label{alphaRich}
\alpha_s(k_t^2)={{12 \pi }\over{(33-2N_f)}}{{p}\over{\ln[1+p({{k_t^2}
\over{\Lambda^2}})^{p}]}}
\end{equation}
which coincides with the usual one-loop expression for large 
 values of $k_\perp$, while going to a singular
 limit for small $k_\perp$. In this model, $\alpha_s$ depends upon 
the parameter set \{$\Lambda,p$\}. 
 The singular expression of eq.(\ref{alphaRich}) is inspired by the
Richardson potential \cite{RICHARDSON} used in quarkonium spectroscopy. 
The Richardson
potential can be connected to a
singular $1/k^2_{\perp}$ behaviour of $\alpha_s$ in the infrared limit,
a singularity which is not dangerous in bound state problems, where the 
Schroedinger equation selects only those solutions for which the momentum
is fixed by the
stability condition. For this problem, and as discussed in \cite{bn},
the expression we have chosen should be considered as a toy model,
in  which the singular behaviour of $\alpha_s$ (if any) can be
modulated through the singularity parameter $p$.
One should also notice that the singular limit of the above equation is 
not an observable.
Phenomenologically,
 one never measures $\alpha_s$ in the $k_\perp\rightarrow 0$
 limit, since this limit corresponds to emission of a very soft gluon,
 in which case summation, and hence integration over $k_\perp$, is
mandatory. In other words, what really matters is the integrability of the 
function, 
since  observable quantitites (soft gluons are observed only as 
overall energy momentum imbalance carried away by soft particles, but not 
measured individually) always involve an integration over the infrared region.
In what follows we shall always use the set $a=2, \Lambda=0.2\ GeV$ for
the frozen $\alpha_s$ model, whereas for the singular case, while we shall vary
the singularity parameter $p$, we shall adopt the value
 $\Lambda=0.1\ GeV$\cite{NAK}.

Let us now examine the function $h(b;M,\Lambda)$. This function does not
allow for a closed form expression, and needs to be numerically evaluated.
Useful analytical approximations can be found in the appendix.

\begin{figure}[htb]
\begin{center}
\mbox{\epsfig{file=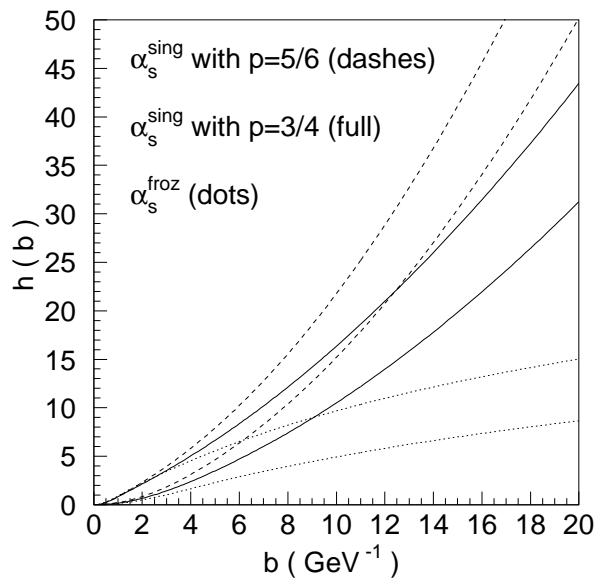,height=6cm}}
\vspace{2cm}
\caption{Comparison between numerically integrated 
 expressions for $h(b,M,\Lambda)$
for singular and frozen $\alpha_s$. $M=1 $ (lower) and $4\ GeV$ (upper).}
\label{hcomp}
\end{center}
\end{figure}

The dependence of the function $h(b;M,\Lambda)$ upon the
infrared behaviour of $\alpha_s$ is shown in
Fig.\ref{hcomp}, where we have plotted in the same graph the exactly integrated
expression for the function $h(b,M,\Lambda)$ for the frozen case, eq.(11),
 and for the
singular case, eq.(12),  for two values of the parameter p. 
For each case, we have evaluated $h(b)$ for the two values $M=1$ and $4\ GeV$,
which correspond to the interesting range $\sqrt{s}=50\div 10^{4}\ GeV$,
for $p_{tmin}=1\div 2\ GeV$ (see Fig. \ref{qmaxfig}). Although at very small
$b$ ($b\le 0.2 \ GeV^{-1}$) the values are not very different,
 at larger $b$-values there is
an increasing discrepancy between the two formulations. The large $b$ 
region below $\approx 10 \ GeV^{-1}$
is the one which matters most for the total cross-section analysis, where 
the figures then show that the infrared behaviour of $\alpha_s$
plays an important role in the rise of the cross-section. In the
next sections we shall study the difference in  $A(b)$ and then in
the number of collisions, given the same jet-cross-section.

\section{The overlap function $A(b)$}

\begin{figure}[htb]
\begin{minipage}{2.4in}
\centering
\leavevmode
\mbox{\epsfig{file=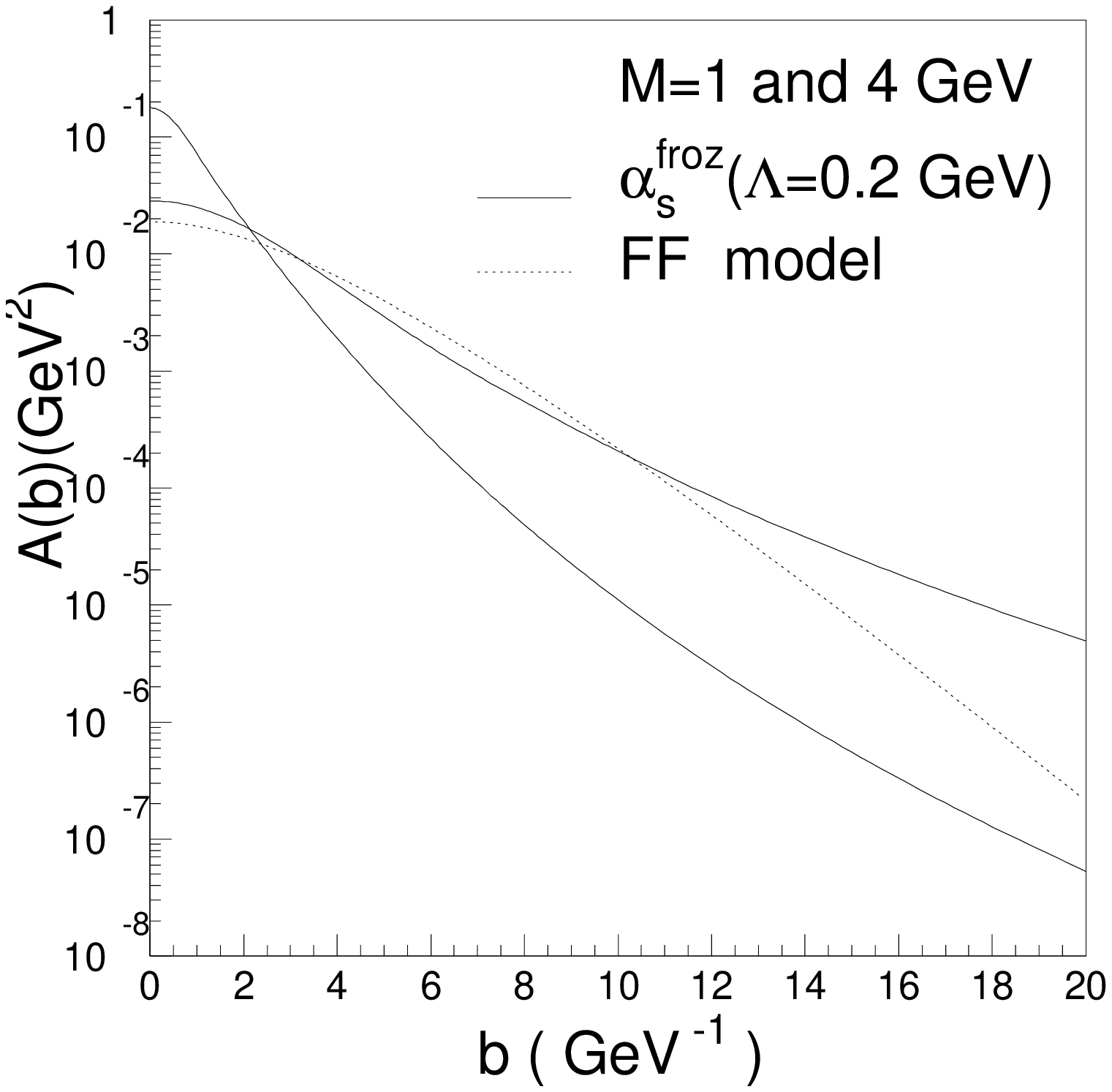,height=6.cm}}
\caption{The overlap function $A_{BN}(b)$ for frozen $\alpha_s$
and comparison with the Form Factor model}
\label{aobf}
\end{minipage}
\hfill
\begin{minipage}{2.4in}
\centering
\leavevmode
\mbox{\epsfig{file=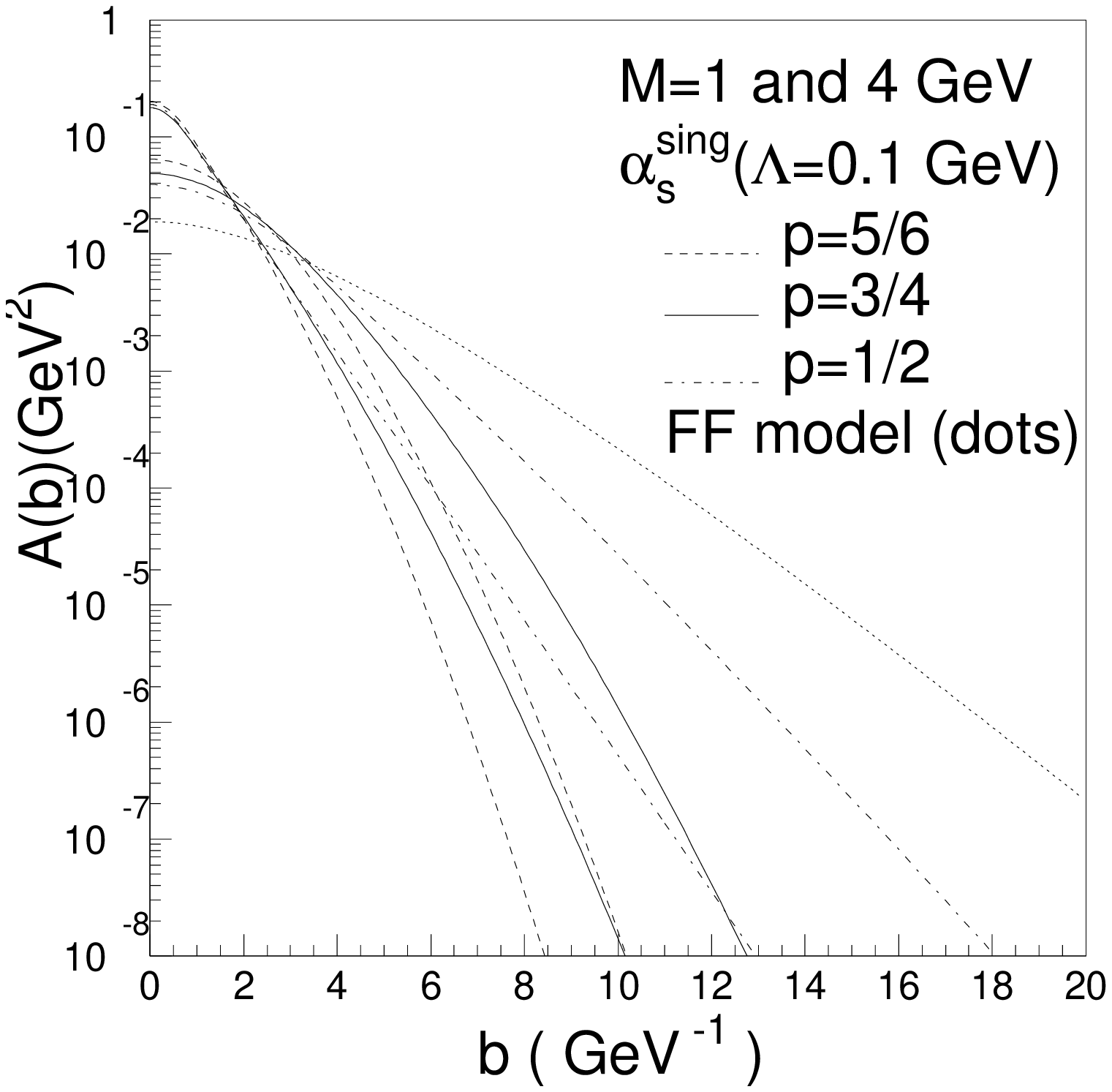,height=6.cm}}
\caption{As in Fig.\protect{\ref{aobf}} for singular $\alpha_s$,
for various values of the parameter $p$.}
\label{aobs}
\end{minipage}
\end{figure}

In this section, we shall calculate numerically  $e^{-h(b;M,\Lambda)}$
and the normalized $A(b)$, for the two cases, frozen and singular
$\alpha_s$. We show in Figs.\ref{aobf},\ref{aobs}
the normalized function $A(b)$ for the frozen 
and  singular $\alpha_s$ possibilities, 
using, for the latter case, three different
values of the parameter $p$ which regulates the singularity.
In both figures we also show the comparison with 
 the function $A(b)$ in the form factor
model, according to which  matter density  in the proton 
is given by the
electromagnetic form factor. With  the usual
 parametrization
\be
{\cal F}_{proton}(q)=\left(
 {{\nu^2}\over{q^2+\nu^2}}\right)^2\ \ \ \ \ \ \ \nu^2=0.71
GeV^2
\ee
the overlap function $A(b)$ in the form factor model has the
expression
\be
A_{FF}(b)={{\nu^2}\over{96 \pi}} \left(\nu b \right)^3 K_3(\nu b)
\ee
In each figure, the various curves correspond to varying
the scale $M$ as described in the previous section, so that they
include a range of energies  $\sqrt{s}=50\div 10^4
\ GeV$ for a range of $p_{tmin}$ between
$1$ and $2\ GeV$.
We see that the frozen $\alpha_s$ case is more
similar to the form factor model, especially at low medium energies,
($50\rightarrow 100\ GeV$), when the proton
is not yet exhibiting the full QCD behaviour.
This is different from  the singular case, where the function
$A(b)$ is always falling with energy more than in the form factor
model. The more singular $\alpha_s$ as $k_t\rightarrow 0$ (larger $p$ values),
the more concentrated at small impact parameter is the overlap function
and hence the less important the large b-values. This will have as 
physical consequence that as the c.m. energy increases, the non-collinearity
of the initial state due to soft gluon emissions will
accordingly increase.
 Clearly this will signify a much more
noticeable effect of soft gluon straggling on the total cross-section.
We shall now see this effect on the average number of
collisions $n(b,s)$.

\section{Average number of collisions}

\begin{figure}[htb]
\begin{minipage}{2.4in}
\centering
\leavevmode
\mbox{\epsfig{file=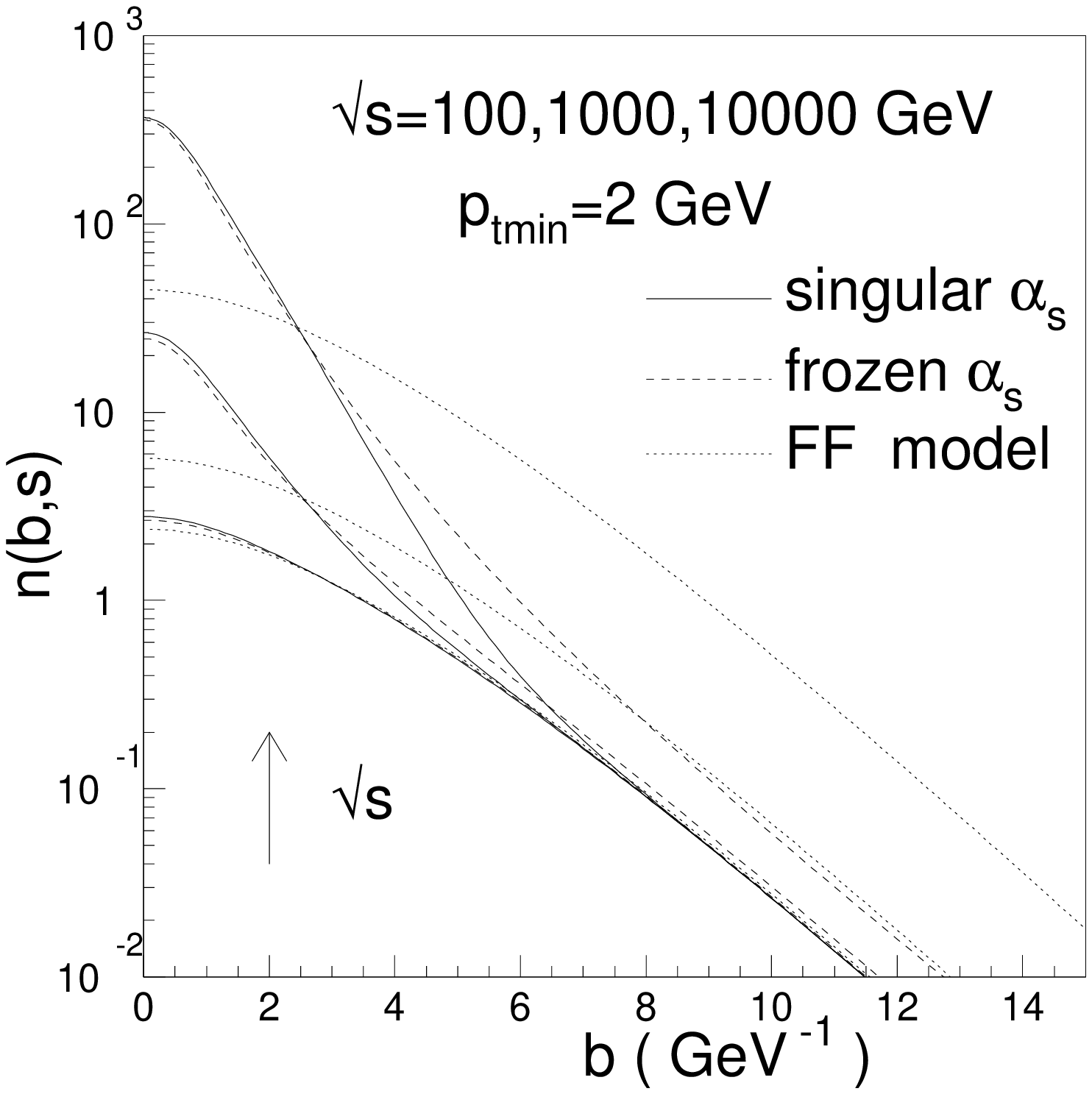,height=6cm}} 
\caption{The average number of collisions for the frozen and singular
$\alpha_s$ case
in comparison with the FF model at various c.m. energy values.}
\label{nbsf}
\end{minipage}
\hfill
\begin{minipage}{2.4in}
\centering
\leavevmode
\mbox{\epsfig{file=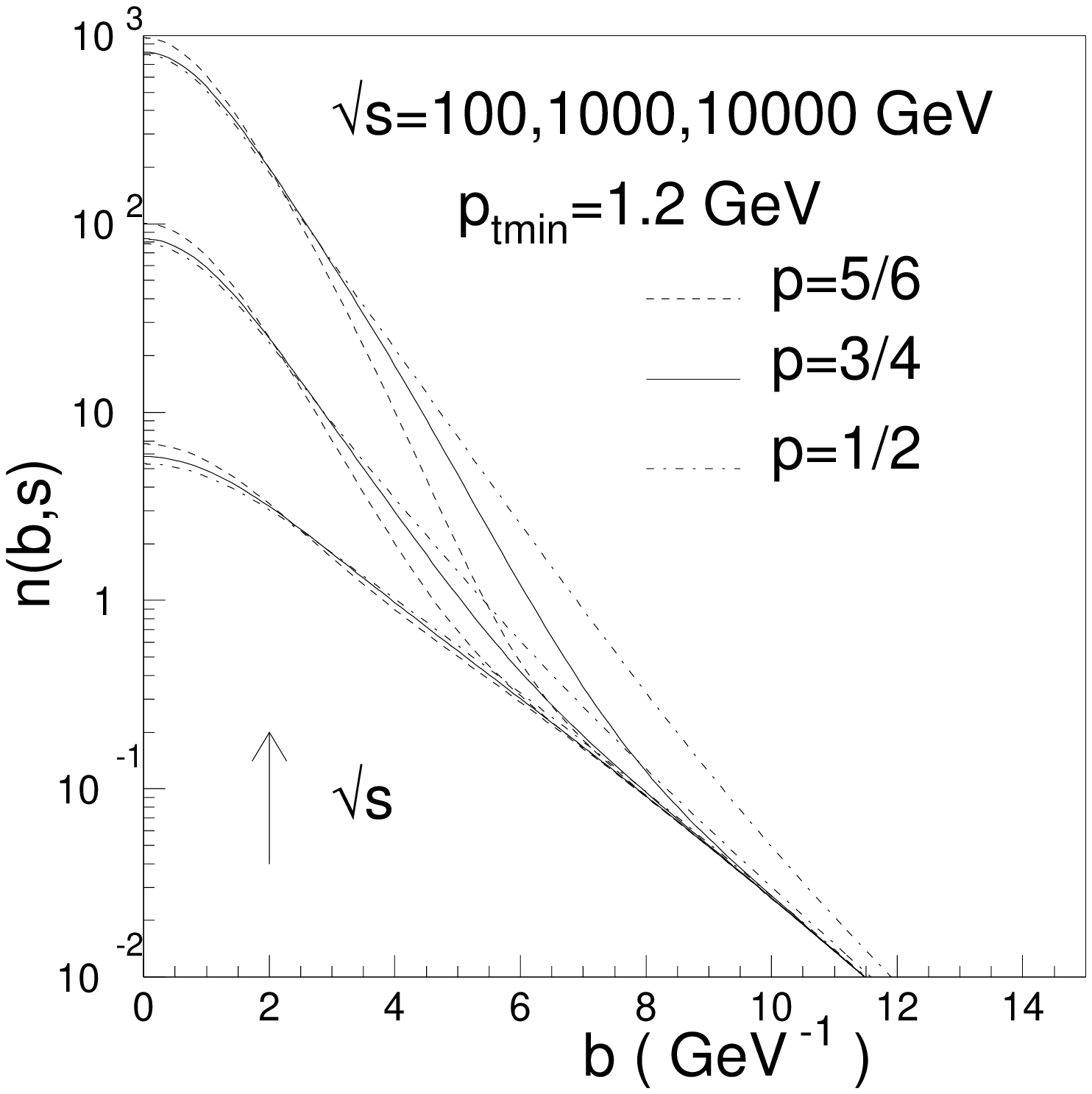,height=6.cm}}
\caption{The average number of collisions for the singular $\alpha_s$ case
for different values of the singularity parameter $p$.}
\label{nbss}
\end{minipage}
\end{figure}

In the eikonalized mini-jet model, the quantity which contains the energy 
dependence of the total cross-section, is the average number of collisions
$n(b,s)$. At low c.m. energy of the colliding particles, this number 
is dominated by contribution from soft, non-perturbative
type events, while  the QCD component, mini-jet like, slowly rises, reaching
a comparable size in the $200\div 300\ GeV$ region. As mentioned in the
first section, one can approximate
the average number of collisions in the entire region as
\be
n(b,s)=n_{soft}(b,s)+n_{hard}(b,s)
\ee
with \be
n_{soft}(b,s)=A_{FF}(b)\sigma_{soft}(s)
\ee
and 
\be n_{hard}(b,s)=A_{BN}(b;M,\Lambda)\sigma_{jet}(s,p_{tmin})
\ee
To study the b-behaviour, we shall introduce the soft term, by using the
 form factor model
for $A(b)$ described in the previous section, which 
 is consistent with a low energy model of the proton in which
only valence quarks play a role in the scattering. In this model all the energy
dependence comes from the cross-section term : we  will parametrize 
$\sigma_{soft}$ so as to reproduce, through the
eikonal,  the low
energy behaviour of the total proton-proton and proton anti-proton
cross-sections.
We found, as best fit to the low energy data with an eikonal formulation
with $n_{hard}=0$
\be
\sigma_{soft}^{pp}=47 +{{46}\over{E^{1.39}}}
\ee
and
\be
\sigma_{soft}^{p{\bar p}}=47 +{{129}\over{E^{0.661}}}+{{357}\over{E^{2.7}}}
\ee
where $E$ is the proton energy in the Laboratory system in GeV and the
 cross-sections are in $mb$.
For $\sigma_{jet}$  we use GRV(LO) densities to
evaluate the proton-proton jet cross-section and 
two different  values of $p_{tmin}=1.2$ and $2\ GeV$, the latter
being the one for which
the total cross-section in the FF model 
passes through the CERN data points at $\sqrt{s}=546\ GeV$. 
In a subsequent section, when we shall try to fit the total 
cross-section data, we shall use other $p_{tmin}$ values.
We can now plot the entire $n(b,s)$ as a function of b, 
for various values
of the center of mass energy $\sqrt{s}$, which corresponds to
various values of the scale M,
as described in the first section. We show this behaviour for the frozen 
and singular
$\alpha_s$ case in Figs.\ref{nbsf}, for $p_{tmin}=2\ GeV$.

\begin{figure}[htb]
\begin{minipage}{2.4in}
\centering
\leavevmode
\mbox{\epsfig{file=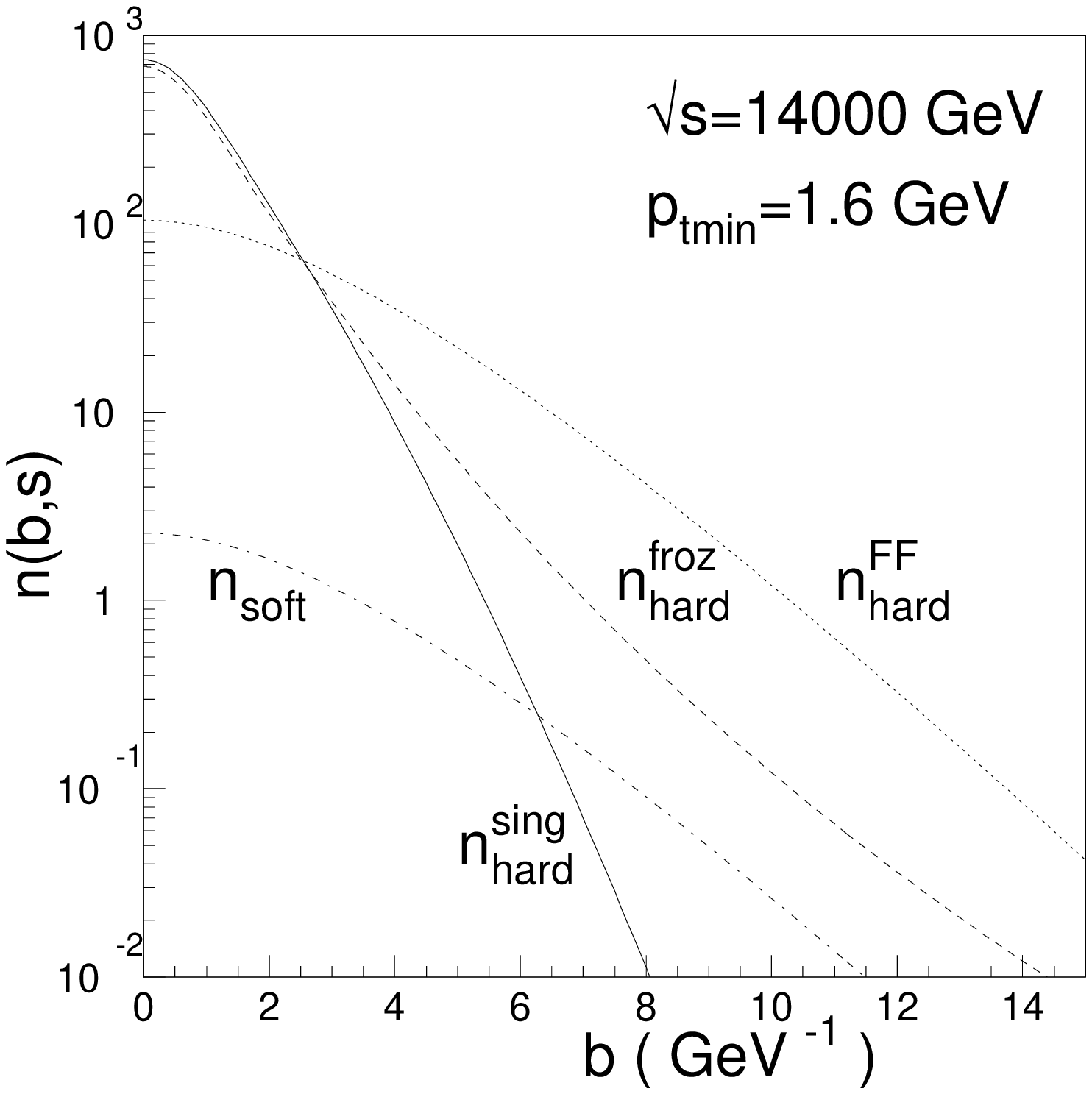,height=6cm}}
\caption{Soft and hard component of n(b,s) in the three models described
in the text.}
\label{nsofthard}
\end{minipage}
\hfill
\begin{minipage}{2.4in}
\centering
\leavevmode
\mbox{\epsfig{file=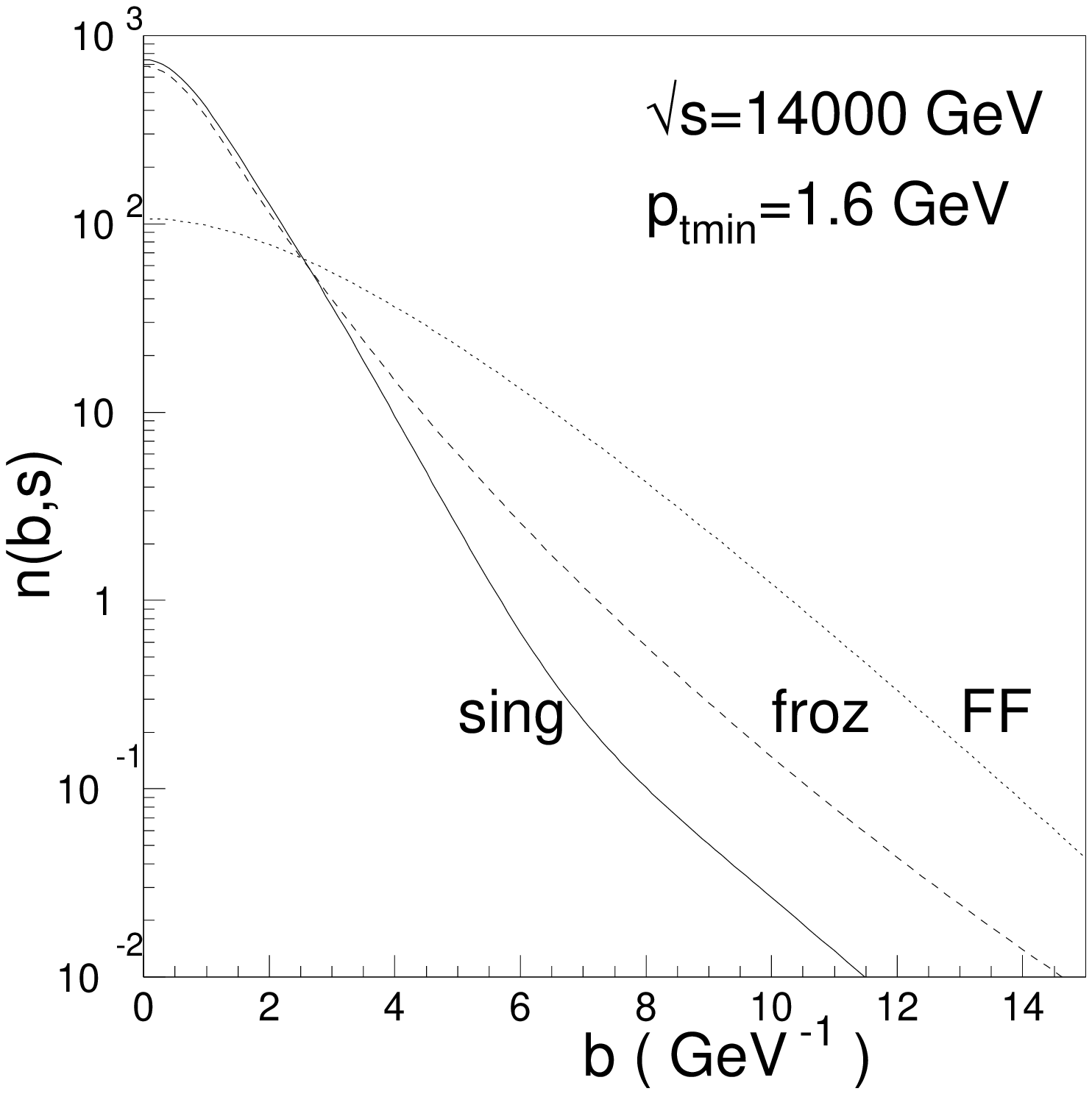,height=6.cm}}
\caption{The average number of collisions in the form factor model and the
Bloch Nordsieck model, at LHC energy.}
\label{3n}
\end{minipage}
\end{figure}
For the frozen $\alpha_s$ model, shown in Fig.\ref{nbsf},
the results are
 compared to a straightforward application of the form 
factor 
model, i.e. with
\be
n_{FF}(b,s)=A_{FF}(b)[\sigma_{soft}+\sigma_{jet}]\ee
We see that at $\sqrt{s}=100\ GeV$, there is still no difference betwen the two
models. On the other hand, as the energy increases, the BN  model
shows a stronger suppression of the large $b$-contribution.
 For the singular case, in order to show variations with
 the singularity parameter $p$, we plot in Fig.\ref{nbss} the result for 
different $p$-values. Notice that the $p$-dependence is related
to the values of $k_t$ probed, i.e. by the $M$ values,which are smaller the 
smaller 
$p_{tmin}$ is. Thus, for
$p_{tmin}=2\ GeV$ for instance, there is very little 
difference among the various
curves, at any given energy. This reflects the fact that
the upper integration limit in eq.(\ref{hdb}) is
a relatively large $M$ ($3\div 4\ GeV$) value, so that
the overall function is not very sensitive to the
infrared region. It should be noted that for smaller $p_{tmin}$ values, 
like the ones actually used for fitting the total cross-sections
 in the next section, the dependence upon $p$ is much more noticeable.
We show one such case in Fig.\ref{nbss}.

The next figure, Fig.\ref{nsofthard},
shows a break down of the average number of collisions into 
the soft and the hard component.
In the present analysis we are
not changing the  soft component, which appears as the dash-dotted curve,
  and the
figure shows, at a given high (LHC) energy, how the hard
part would be different in the three models, i.e.
in general more  peaked at small b for the Bloch-Nordsieck model,
and  in particular falling faster the stronger the singularity of $\alpha_s$.
  At lower energies, where the mini-jet contribution is
less important,  these discrepancies would be much reduced. 
So, in this picture, while keeping a similar  b-distribution at low
energy, we  quantitatively enhance  small
b-collisions at high energy, though QCD soft gluon emission.
 The change in
the $b$-distribution introduced in the hard
component by the different models for $A(b)$ 
is responsible for the changed shape of $n(b,s)$ between the
form factor and the other two models. 
  The direct comparison among the three 
models is shown in Fig.(\ref{3n}) where the average number of
collisions at $\sqrt{s}=14\ TeV$ is plotted for a choice of
the various parameters as indicated.
Apart from the change in shape, it can be noticed that
 the frozen
$\alpha_s$ case corresponds to a behaviour intermediate between
the form factor model, and the singular $\alpha_s$ case.
We also see from this figure that the
 range of values of the $b-$parameter most important to the total 
cross-section calculation
changes 
in the different models. 

\section{Total cross-sections}

\begin{figure}[htb]
\begin{center}
\mbox{\epsfig{file=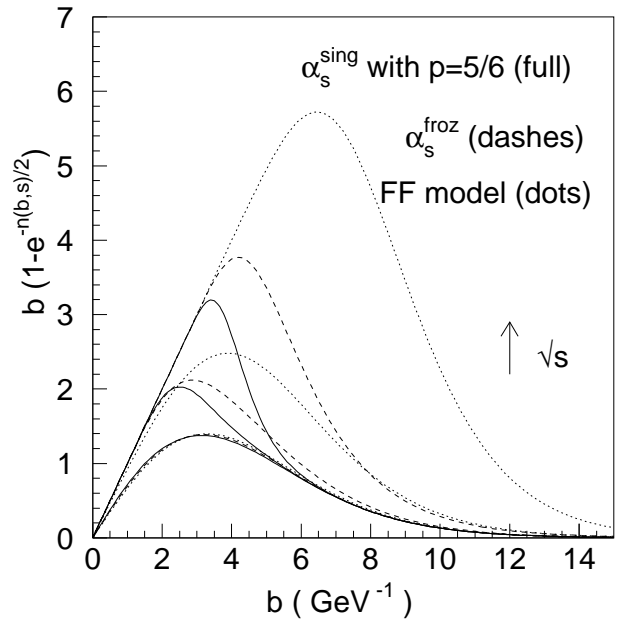,height=10cm}}
\caption{The integrand of the eikonal formulation for $\sigma_{tot}$,
for  $p_{tmin}=2\ GeV$ in the three different models described in the text,
for a range of c.m. energy values 100,1000,10000 GeV}.
\label{int}
\end{center}
\end{figure}

Before attempting the last  relevant phenomenological exercise for the 
calculation of the total proton-proton and proton-antiproton cross-section,
we shall first show how the integrand in eq.(\ref{sigtot}) changes with
energy and which values of $b$ are most relevant for
the calculation of the total cross-section in the various models 
for $A(b)$ we have
just described.
We must stress that this is not an optimization of the many 
parameters from which this model depends : rather an exercise to
show how the Bloch-Nordsieck model for the
impact parameter distribution affects the total cross-section behaviour
in the eikonalized minijet model and how the behaviour of
$\alpha_s$ in the infrared region is related to the rise of the
total cross-section. This is done in Fig.\ref{int}.
The figure shows how much the integrand of eq.(\ref{sigtot}) is peaked at 
different $b$-values as the energy increases, but also as the model for
$A(b)$ changes. And it indicates that the rise with energy of the area under 
the curve,
i.e. the
cross-section, at the same energy shrinks for
the more singular $\alpha_s$ behaviour.

\begin{figure}[htb]
\begin{center}
\mbox{\epsfig{file=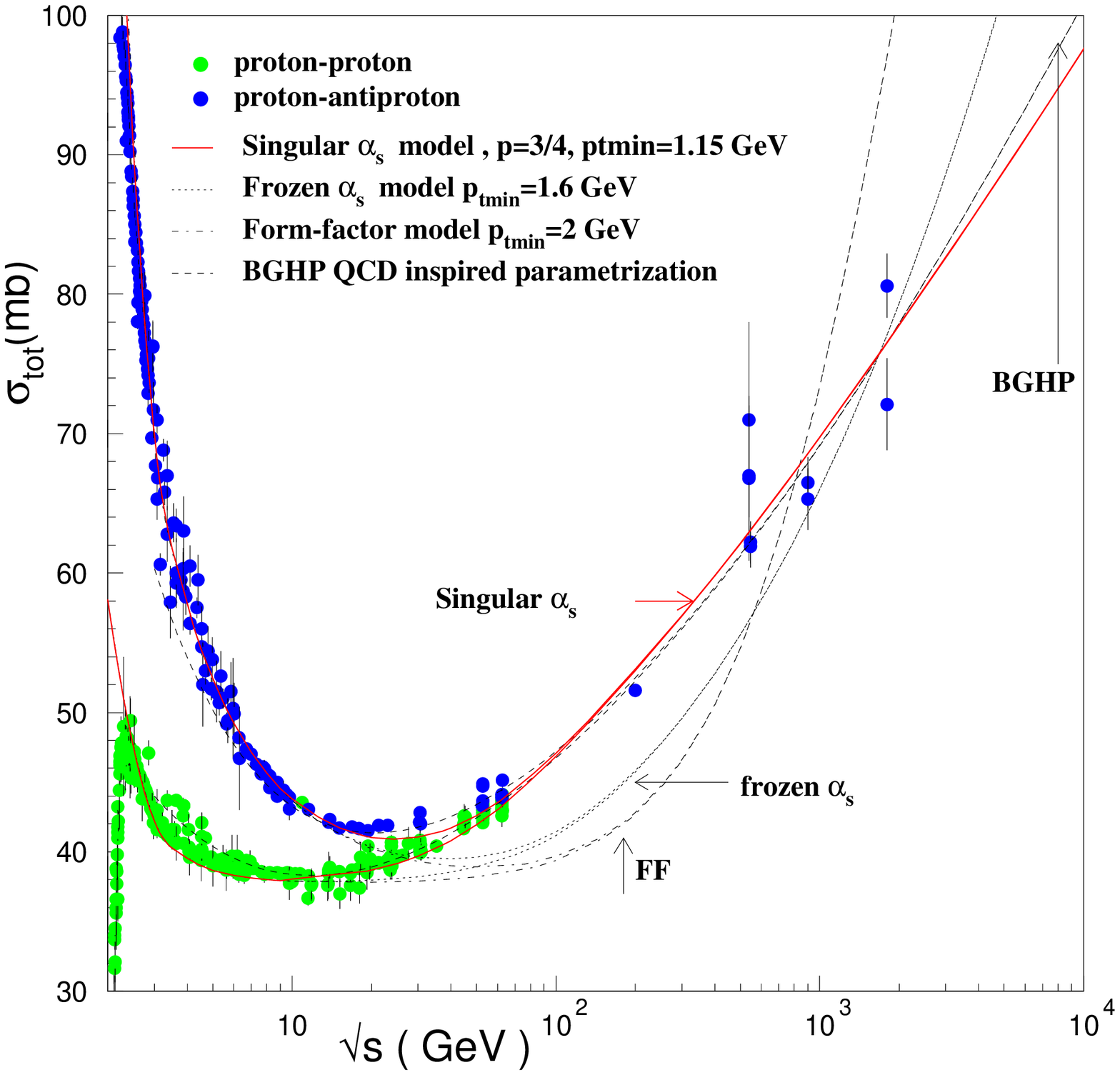,height=10cm}}
\vspace{1cm}
\caption{Total $p$-$p$ and ${\bar p}p$ cross-sections and comparison with 
various models}
\label{sigtotmodels}
\end{center}
\end{figure}

Finally in  Fig.\ref{sigtotmodels} we show the comparison of this
model with proton-proton and proton-antiproton data.
For proton-proton, we only show data up to ISR energies, since
the existing data points in the TeV range are extrapolations from cosmic ray
data \cite{COSMIC}  from $p-air$ collisions
and are partly model dependent\cite{EGLS}.
 For the proton-antiproton data,
we have plotted all the data points published so far from the CERN
$Sp{\bar p}S$ \cite{CERN} and FNAL \cite{FNAL} experiments. This 
introduces a larger band of uncertainty that it is usually shown,
 but  the
purpose of this paper is to indicate the potentiality 
of the Bloch-Nordsieck model rather than to do a best parameter fit, 
and we have opted for a comparison of our results with the
full experimental picture.

We have  studied three different
 formulations of  the eikonal mini-jet model, one for the form factor
model and two for the soft gluon summation model. To choose the
parameters of the mini-jet description, we have selected those
$p_{tmin}$ values which would ensure that the curve can reach the high
 energy points :  for the form factor model this can be accomplished with
 $p_{tmin}=2 \ GeV$ but, as often stressed, with such value
it is not possible  to fit the early
rise of the data. A lower value of $p_{tmin}$ would on the other
hand give curves which rise too much at higher energy and miss the points.
Going to the soft gluon summation model, it must be noticed that
since this model has an energy dependence in the b-behaviour in addition to
the one in the jet-cross-section (common to all the models), 
one can expect that a smaller $p_{tmin}$ could be used, thus
allowing for the earlier rise. In fact, the high energy data, for the
frozen $\alpha_s$ model, can be met with $p_{tmin}=1.6 \ GeV$. Although
with this value, the cross-section starts rising sooner than in the form factor
model, still it is impossible to fit both the early rise as well as the
high energy points. This model depends not only upon $p_{tmin}$ value, but
also on the scale $a$ which regulates the infrared behaviour of 
$\alpha_s$ : the smaller $a$, the more singular the behaviour and the
easier to fit the early rise. Finally, we show the
results for the singular $\alpha_s$ case, with a particular choice of the
parameter $p$ which regulates the  singularity of $\alpha_s$. We can
choose now a rather small value of $p_{tmin}$ to reproduce the early rise,
since at higher energy the increased soft gluon emission reduces drastically
the large-b contribution to the cross-section and does not let it rise as
much as in the other models.
Our results are compared with a multiparameter
fit from a QCD inspired model\cite{BLOCK},
 which 
has recently been used to successfully predict photon photon total 
cross-sections\cite{BGHP}.   These results are not
very different from
the ones obtained using the Regge-Pomeron exchange picture
\cite{DL}, but the model in \cite{BGHP} is closer in spirit to 
the one discussed here,
with the energy rise due to 
 the rise of the QCD jet cross-section.

The results of this figure shows that it is possible to
have a rise in agreement both with the intermediate energy data as well as with
the Tevatron data : this result is obtained using a single eikonal
function, usual QCD parton densities and minijet cross-sections with $p_{tmin}$
in the $1\div 2\ GeV$ range. To follow the beginning of the rise, one needs a
rather low $p_{tmin}$. In general such low values imply too fast a growth
of the total cross-section, in our case this fast growth is tampered by the
 increasing number of soft gluon emission phenomena at small $k_t$.

\section{Conclusions}

We have presented a detailed numerical analysis of a Bloch-Nordsieck approach
to
the impact parameter distribution of partons in the context of the
eikonal mini-jet model for total hadronic cross-sections. We have shown 
that the proposed soft gluon summation expression plays an important role in 
softening the rise of the cross-section due to mini-jets and have
studied the role which the infrared behaviour
of $\alpha_s$ plays in it.
\section{Acknowledgments}
This work was supported in part by EEC-TMR Contract N.CT9800169 and by CICYT 
Contract N. AEN 96-1672.  

\appendix
\section{Approximate Expressions for $h(b,M,\Lambda)$}
\label{appenda}
\newcounter{zahler}
\renewcommand{\theequation}{\Alph{zahler}.\arabic{equation}}
\setcounter{zahler}{1}
\setcounter{equation}{0}
In this section, we show some analytic approximations to the
 function $h(b,M,\Lambda)$.
Here we shall restrict our attention to values of M relevant to
 the total cross-section calculations, i.e. values in the few GeV range.
Since as the total c.m. energy
increases, M increases from 0.5 to 4 GeV, the region of $b>1/M$ ($b<1/M$) 
 corresponds to values of $b$ larger (smaller) than 
 $2.5\ GeV^{-1}$, at low $\sqrt{s}$, 
down to  
 $0.2\ GeV^{-1}$ for the highest $\sqrt{s}$ values. 
In other words, in the integration, small and large b-values are an
 energy dependent concept :
at very small $\sqrt{s}$, small b, i.e. $b<1/M$ means values of
$b$ less than $2.5\ GeV^{-1}$, 
whereas at very high energy large b-values mean $b> 0.2\ GeV$.
We shall now start studying $h(b;M\Lambda)$ in the frozen $\alpha_s$ case,
and distinguish three cases :\begin{enumerate}
\item{} $bM<1$
\item{} $bM>1$, $ba\Lambda<1$
\item{} $bM>1$, $ba\Lambda>1$
\end{enumerate}

In order to obtain a closed form expression to better study
the function, we shall adopt the following approximations :
\begin{description}
\item{} $\alpha_s(k<a\Lambda)\equiv {\bar \alpha_s}= 
 \sd {{12 \pi}\over{27 ln(a^2)}}$
\item{}  $\alpha_s(k>a\Lambda)= 
\sd {{12 \pi}\over{27 ln{{k^2}\over{\Lambda^2}}}}$
\item{} $\ln {{M+\sqrt{M^2-k^2}}\over{M-\sqrt{M^2-k^2}}}
\approx \sd 2\ln{{2M}\over{k}}\ \ \ \ k\approx 0$ and
\item{} $\ln {{M+\sqrt{M^2-k^2}}\over{M-\sqrt{M^2-k^2}}}
\approx \sd 2\ln{{M}\over{k}}\ \ \ \ $ for $k$ values not
 in the infrared region.
\item{} $1-J_0(x)=\sd {{x^2}\over{4}}\ \ \ \ \ x<1$
\item{} $1-J_0(x)=1 \ \ \ \ \ x>1$
\end{description}
Then, one can break the integral from $0\rightarrow M$ into various intervals
in which one can approximate the integrand and perform the
integration. According to the three cases indicated above, one then 
obtains the following approximate expression :
\begin{eqnarray}
\label{halphafroz1}
  bM&<& 1 \\
h(b,M,\Lambda)&=&{{2c_F}\over{\pi}}
\left[{\bar \alpha_s}{{b^2}\over{2}}\int_0^{a\Lambda}k dk \ln {{2M}\over{k}}
+{\bar b} {{b^2}\over{4}}\int_{a\Lambda}^M k dk 
{{\ln {{M}\over{k}}}\over{\ln{{k}\over{\Lambda}}}}\right] \nonumber \\
 &=&{{2c_F}\over{\pi}}\Biggl \{ {\bar \alpha_s}{{b^2 \Lambda^2 a^2}\over{8}}
\left[1+2\ln{{2M}\over{a\Lambda}}
\right]+\nonumber \\
  &+ &{\bar b}{{b^2M^2}\over{8}}\left\{ {{a^2\Lambda^2}\over{M^2}}-1
+2{{\Lambda^2}\over{M^2}}\ln{{M}\over{\Lambda}}\left[
 li\left({{M^2}\over{\Lambda^2}}\right)-li(a^2)\right]\right\} \Biggr \}
\nonumber \end{eqnarray}
For $bM>1$, one distinguishes between two cases : $b$ larger or
smaller than $1/a\Lambda$, so that the integral can now be divided as
follows :
\begin{eqnarray}
\label{halphafroz2}
for\ &{{1}\over{M}}&< b <{{1}\over{a\Lambda}} \\
h(b,M,\Lambda)&=&{{2c_F}\over{\pi}}
\left[{\bar \alpha_s}{{b^2}\over{2}}\int_0^{a\Lambda}k dk \ln {{2M}\over{k}}
+{\bar b} {{b^2}\over{4}}\int_{a\Lambda}^{{1}\over{b}} k dk 
{{\ln {{M}\over{k}}}\over{\ln{{k}\over{\Lambda}}}}
+{\bar b}\int_{{1}\over{b}}^M {{dk}\over{k}}
{{\ln{{M}\over{k}}}\over{\ln{{k}\over{\Lambda}}}}\right] \nonumber \\
 &=&{{2c_F}\over{\pi}}\Biggl [
{\bar \alpha_s}{{b^2 \Lambda^2 a^2}\over{8}}\left[1+2\ln{{2M}\over{a\Lambda}}
\right ]+\nonumber \\
  &\ &{\bar b}{{b^2\Lambda^2}\over{8}}\left\{ a^2-{{1}\over{b^2\Lambda^2}}
+2\ln{{M}\over{\Lambda}}\left[
 li\left({{1}\over{b^2\Lambda^2}}\right)-li(a^2)\right]\right\}+\nonumber \\
 &\ &{\bar b}\left[ \ln{{M}\over{\Lambda}}\ln{{\ln{{M}\over{\Lambda}}}\over
{\ln{{1}\over{b\Lambda}}}}-\ln(Mb)\right] \Biggr ] \nonumber
\end{eqnarray}
or as
\begin{eqnarray}
\label{halphafroz3}
for \ &{{1}\over{M}}& <{{1}\over{a\Lambda}}< b \\ 
h(b,M,\Lambda) &=&{{2c_F}\over{\pi}}
\left[{\bar \alpha_s}{{b^2}\over{2}}\int_0^{{{1}\over{b}}}k dk
 \ln {{2M}\over{k}}
+2 {\bar \alpha_s} \int^{a\Lambda}_{{1}\over{b}} {{dk }\over{k}}
\ln {{M}\over{k}}
+{\bar b}\int_{a\Lambda}^M {{dk}\over{k}}
{{\ln{{M}\over{k}}}\over{\ln{{k}\over{\Lambda}}}}\right] \nonumber \\
 &=&{{2c_F}\over{\pi}}\Biggl [
{{{\bar \alpha_s}}\over{8}}\left[1+2\ln(2Mb)\right]+\nonumber \\
  &\ & 2{\bar \alpha_s}\left\{ \ln(Mb)\ln(a{\Lambda}b)
-{{1}\over{2}}\ln^2{(a \Lambda b)}\right\}+\nonumber \\
 &\ &{\bar b}\left[ \ln{{M}\over{\Lambda}}\ln {{\ln{{M}\over{\Lambda}}}
\over{\ln a }}-\ln {{M}\over{a\Lambda}}\right] \Biggr ]\nonumber
\end{eqnarray}

The last decomposition is the one to use to study the
large  b limit, whereas the first one corresponds to the small b limit.

\begin{figure}[htb]
\begin{minipage}{2.40in}
\centering
\leavevmode
\mbox{\epsfig{file=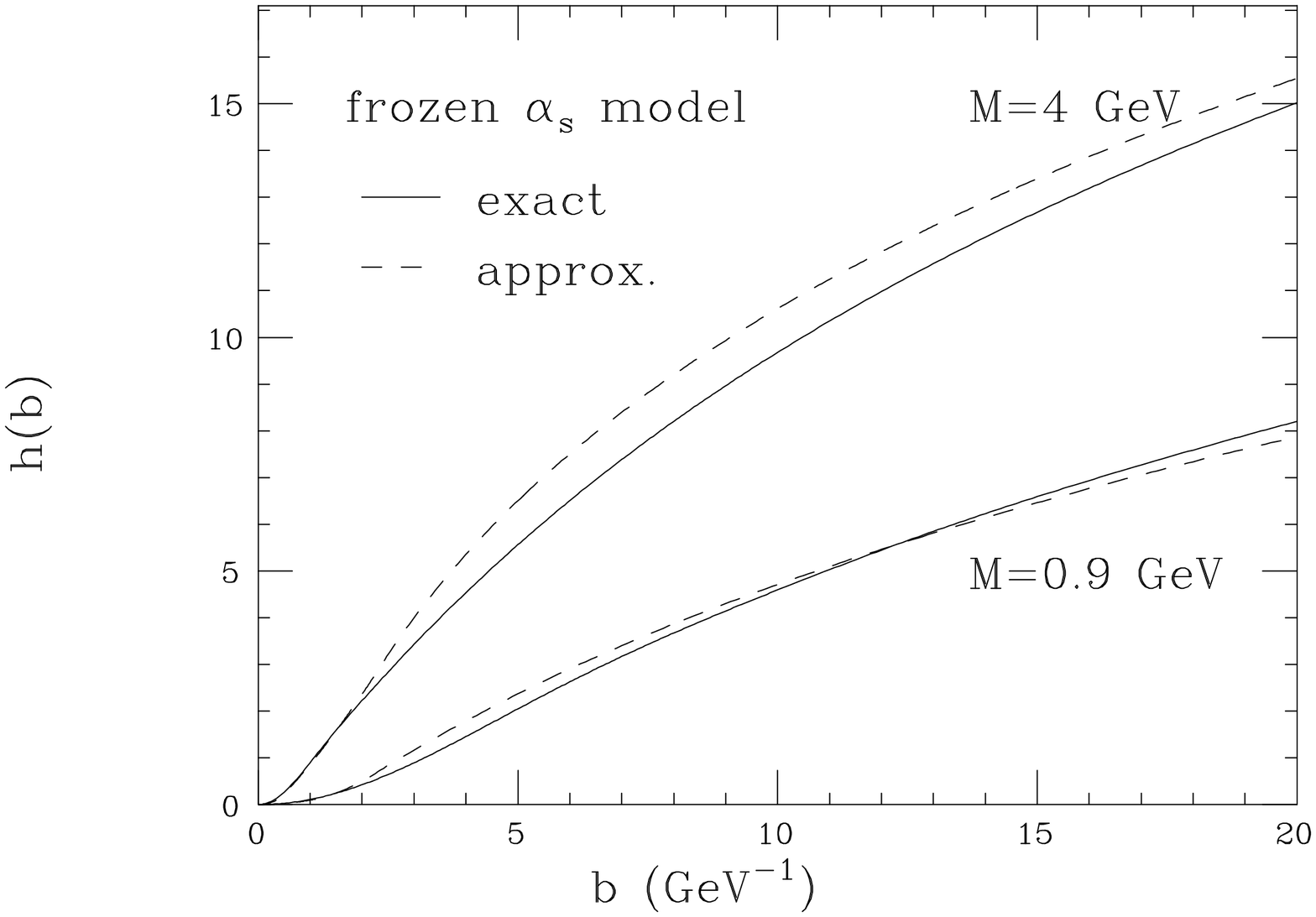,height=6cm,angle=90}}
\caption{Comparison between the approximate and the actual numerical
integration for $h(b,M,\Lambda)$ for various values for M, in
the frozen $\alpha_s$ model}
\label{halfroz}
\end{minipage} 
\hfill
\begin{minipage}{2.4in}
\centering
\leavevmode
\mbox{\epsfig{file=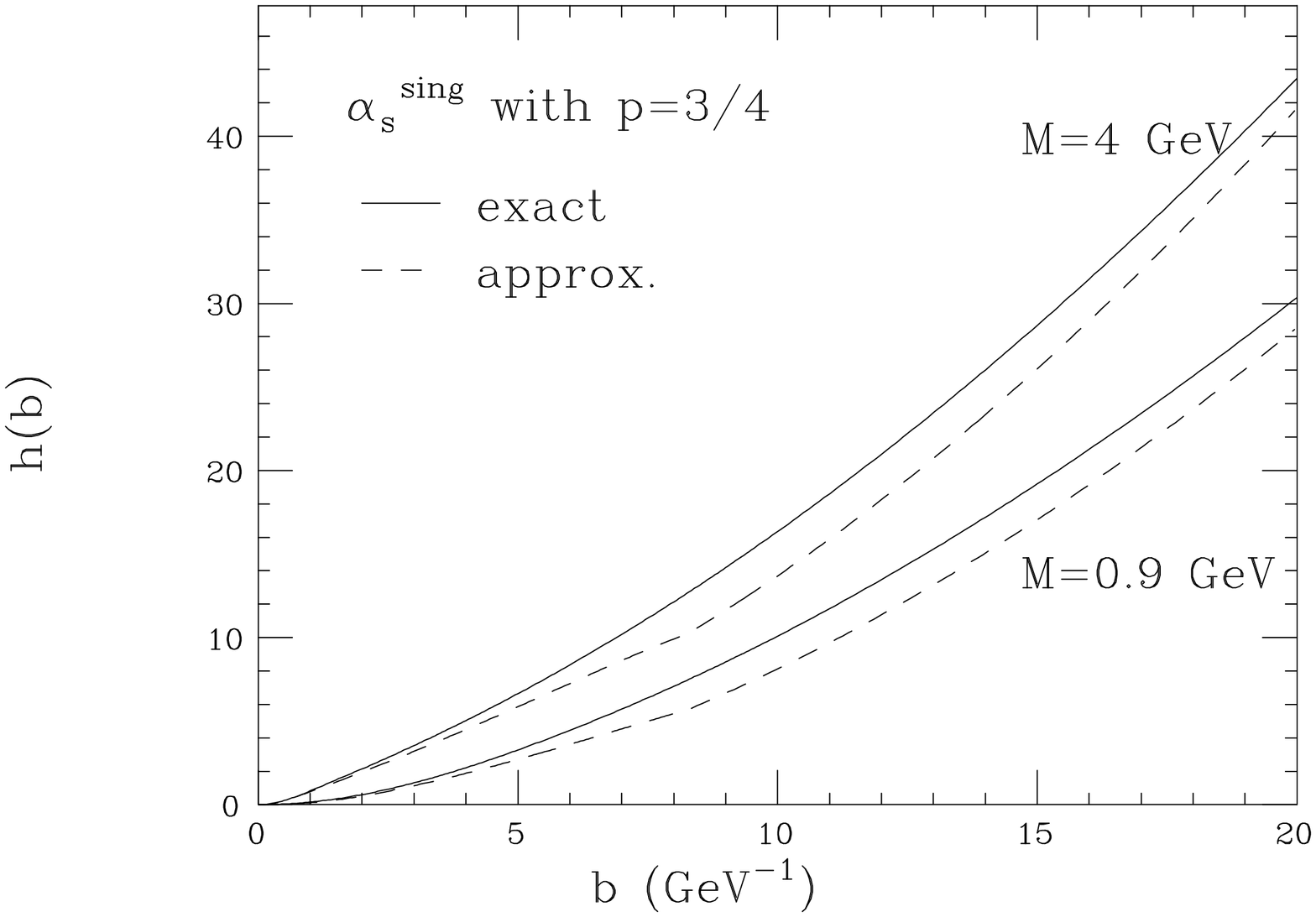,height=6cm,angle=90}}
\caption{Comparison between the approximate and the actual numerically
computed expression for $h(b,M,\Lambda)$ for
the singular $\alpha_s$ model}
\label{hals}
\end{minipage}
\end{figure}

For the singular $\alpha_s$ case we adopt similar approximations, except 
that now
\begin{description}
\item{} 
$\alpha_s(k<N_p\Lambda)=\sd {\bar b} \left({{\Lambda}\over{k}}\right)^{2p}$
\item{} 
$\alpha_s(k>N_p\Lambda)=\sd {{{\bar b}}\over{\ln({{k^2}\over{\Lambda^2}})}}$
\end{description}
where $N_p=\sd \left({{1}\over{p}}\right)^{1/2p}$ is a number of order unity.
 For 
p=1/2, indeed
$N_p=2$ and the two regions, small and large k, coincide with
 those  in the frozen $\alpha_s$ case
with a=2.
The expressions for $h(b;M,\Lambda)$ in this case become :
\begin{eqnarray}
\label{halphas1}
b & < &{{1}\over{M}} \\
h(b;M,\Lambda) &=&{{2c_F}\over{\pi}}\left[{\bar b}{{b^2}\over{2}}
(\Lambda)^{2p}\int_0^{\Lambda
N_p} {{dk}\over{k^{2p-1}}}\ln{{2M}\over{k}}+{\bar b}{{b^2}\over{4}}
\int_{\Lambda N_p}^M k dk {\ln{{M}\over{k}}\over{\ln{{k}\over{\Lambda}}}}
\right ] \nonumber \\
 &=&{{2c_F}\over{\pi}}\Biggl [ 
{{{\bar b}(N_p^2)^{1-p}}\over{8(1-p)}} b^2 \Lambda^2
\left( 2\ln{{2M}\over{\Lambda N_p}}+{{1}\over{1-p}}\right)\nonumber +\\
 &\ &{{{\bar b}}\over{8}}b^2M^2\left\{ {{N^2_p\Lambda^2}\over{M^2}}-1 + 
{{\Lambda^2}\over{M^2}} 2 \ln{{M}\over{\Lambda}}\left[ 
li\left({{M^2}\over{\Lambda^2}}\right )-li(N_p^2)\right] \right\} \Biggr ]
\nonumber \end{eqnarray}
and for the $bM>1$ case one will have the two possibilities,
\begin{eqnarray}
\label{halphas2}
for\ {{1}\over{M}}&<&b
<{{1}\over{N_p\Lambda}}\\
h(b,M,\Lambda) &=&{{2c_F}\over{\pi}} \left[ {\bar b} {{b^2}\over{2}}\Lambda^{2p}
\int_0^{N_p\Lambda} {{dk}\over{k^{2p-1}}} \ln {{2M}\over{k}}+
{\bar b}{{b^2}\over{4}}\int_{N_p\Lambda}^{{{1}\over{b}}}kdk {{\ln{{M}\over{k}}
}\over{\ln{{k}\over{\Lambda}}}}+{\bar b}\int_{{1}\over{b}}^M {{dk}\over{k}}
{{\ln{{M}\over{k}}}\over{\ln{{k}\over{\Lambda}}}} \right] \nonumber \\
 &= &{{2c_F}\over{\pi}}\Biggl [ {\bar b} {{b^2\Lambda^2}\over{8}}
{{(N_p^2)^{1-p}}\over{1-p}} \left( 2\ln{{2M}\over{N_p\Lambda}}+{{1}\over{1-p}}
\right) +\nonumber \\
&\ &{{{\bar b}}\over{8}}\left\{ N^2_pb^2\Lambda^2 -1 +2 \ln{{M}\over{\Lambda}}
b^2\Lambda^2 \left[ li({{1}\over{b^2\Lambda^2}})-li(N_p^2)\right] \right\}+
\nonumber \\
 &\ & {\bar b} \left[\ln{{M}\over{\Lambda}}
\ln {{ \ln{{M}\over{\Lambda}}}\over{\ln{{1}\over{b\Lambda}}}}-\ln(Mb)\right]
 \Biggr]\nonumber
\end{eqnarray}
and the other case
\begin{eqnarray}
\label{halphas3}
b>{{1}\over{N_p\Lambda}}&>&{{1}\over{M}}\\
h(b,M,\Lambda) &=&{{2c_F}\over{\pi}}\left[ {\bar b} {{b^2\Lambda^{2p}}\over{2}}
\int_0^{{1}\over{b}}{{dk}\over{k^{2p-1}}} \ln {{2M}\over{k}}+
2 {\bar b} \Lambda^{2p}\int_{{1}\over{b}}^{N_p\Lambda} {{dk}\over{k^{2p+1}}}
\ln {{M}\over{k}}+{\bar b} \int_{N_p\Lambda}^M {{dk}\over{k}} 
{{\ln{{M}\over{k}}}\over{\ln {{k}\over{\Lambda}}}}\right] \nonumber \\
 &=&{{2c_F}\over{\pi}} \Biggl [ {{{\bar b}}\over{8(1-p)}} (b^2\Lambda^2)^p
\left[ 2\ln(2Mb)+{{1}\over{1-p}}\right] +\nonumber \\
 &\ &{{\bar b}\over{2p}}(b^2\Lambda^2)^p \left[2\ln(Mb)-{{1}\over{p}}\right]
+{{\bar b}\over{2pN_p^{2p}}}\left[-2\ln{{M}\over{\Lambda N_p}}+{{1}\over{p}}
\right] + \nonumber \\
 &\ & {\bar b} \ln {{M}\over{\Lambda}}\left[\ln {{\ln{{M}\over{\Lambda}}}\over
{\ln{N_p}}}-1+{{\ln{N_p}}\over{\ln{{M}\over{\Lambda}}}} \right] \Biggr ]
\nonumber
\end{eqnarray}
This approximation is reasonably accurate, as one
can see from Figs.\ref{halfroz}-\ref{hals}, where we have plotted both
 the approximate 
and the exact
expressions from the above equations for the two different models
for $\alpha_s$.

\end{document}